\documentclass[lettersize,journal]{IEEEtran}
\usepackage{amsmath,amsfonts,bm}
\usepackage{algorithmic}
\usepackage{algorithm}
\usepackage{array}
\usepackage[caption=false,font=footnotesize]{subfig}
\usepackage{textcomp}
\usepackage{stfloats}
\usepackage{url}
\usepackage{verbatim}
\usepackage{graphicx}
\usepackage{cite}
\usepackage{tikz}
\usepackage[background=none, arrow=red]{callouts}
\usepackage{multirow}
\usetikzlibrary{matrix,babel, arrows.meta, calc,positioning, shapes.misc,graphs,quotes, decorations.pathmorphing, decorations.pathreplacing, decorations.shapes}
\DeclareMathAlphabet{\mathpzc}{OT1}{pzc}{m}{it}

\begin{document}

\title{Wideband Sample Rate Converter Using Cascaded Parallel-serial Structure for Synthetic Instrumentation}

\author{Ruiyuan~Ming, Peng~Ye, Kuojun~Yang,~\IEEEmembership{Member,~IEEE,}, Zhixiang~Pan, Li~chen, Xuetao~Liu%
\thanks{P. Ye is with the Faculty of the University of Electronic Science and Technology of China.}
}

%

\maketitle

\begin{abstract}
A sample rate converter(SRC) is designed to adjust the sampling rate of digital signals flexibly for different application requirements in the broadband signal processing system. In this paper, a novel parallel-serial structure is proposed to improve the bandwidth and flexibility of SRC. The core of this structure is a parallel decimation filter followed by a serial counterpart, the parallel part is designed to process high sampling rate data streams, and the serial part provides high flexibility in decimation factor configuration. A typical combination of cascaded integral comb filter(CIC) and halfband filter is utilized in this structure, the serial recursive loop which limits the processing ability of the CIC filter is transformed into a parallel-pipeline recursive structure. In addition, the symmetry property and zero coefficient of the halfband filter are exploited with the polyphase filter structure to reduce resource utilization and design complexity. In the meantime, the decimation factor of the CIC filter can be adjusted flexibly in a wide range, which is used to improve the system configuration flexibility. This parallel-serial SRC structure was implemented on Xilinx KU115 series field programmable gate array(FPGA), and then applied in a synthetic instrument system. The experiment results demonstrate that the proposed scheme significantly improves the performance of SRC in bandwidth and flexibility.
\end{abstract}

\begin{IEEEkeywords}
Sample rate converter, parallel-serial structure, CIC filter,  halfband filter, FPGA, synthetic instrumentation.
\end{IEEEkeywords}

\section{Introduction}
\IEEEPARstart{T}{he} interconnection of different digital signal processing systems operating at different sampling rates makes SRC a fundamental operation in most modern signal processing chains \cite{a1,a2,a3}. From a design point of view, SRC can be inferred either in the time or frequency domain \cite{b1}. Compared with the frequency domain method, the time domain method has lower complexity, which makes it a better candidate for industry application. From an implementation point of view, there are several applicable structures for SRC, including farrow, polyphase, cascaded CIC-halfband structure and so on\cite {c1,c2,c3,CIC_Halfband}. A Farrow structure is used to implement fractional sampling rate conversion. The polyphase structure can process high-speed parallel data streams, but needs too many multiplier resources \cite{d1,d2}. The cascaded CIC-halfband structure consumes fewer multipliers, however, it is designed to process single-lane data. To take full advantage of these implementation structures and overcome the corresponding limits, an improved structure is proposed in this paper, in which a cascaded CIC-halfband structure is converted to a parallel form. On one hand, the ability to process high-speed data streams has been significantly improved. On the other hand, resource utilization has been reduced to an acceptable level.
 
\begin{figure*}[!t]
	\centering
	\includegraphics{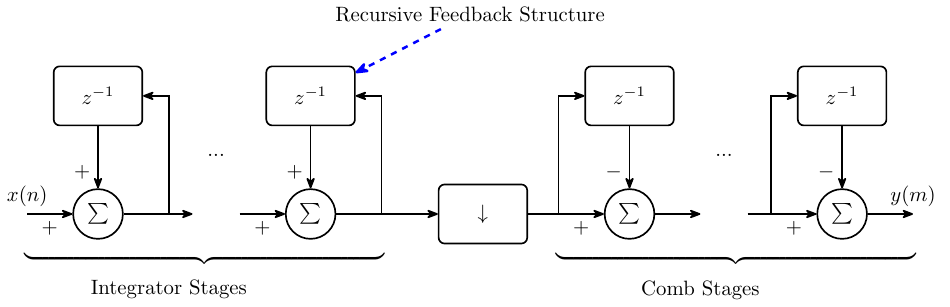}
	\caption{Structure of cascaded decimation integrator comb filter. It is composed of an integrator, downsampler, and comb filter. The feedback structure in the integral part limits the processing rate of the CIC filter.}
	\label{fig CIC}
\end{figure*}

The CIC filter was first introduced by Hogenaur for decimation and interpolation in digital signal processing system.\cite{cic}, it is commonly employed when a large downsampling ratio is required, because of its reasonable performance and low hardware complexity\cite{f1}. However, the serial recursive structure of the CIC limits its ability to handle high-speed data streams. The structure of CIC is shown in Fig. \ref{fig CIC}, it is composed of the integrator, a downsampler, and a comb filter. The integrator part is a single-cycle recursive structure, Only one addition operation can be done per clock cycle, which greatly limits the processing rate of the CIC.

Typically, there are two methods for improving the processing efficiency of the CIC filter. The first method is to reduce the operation delay of the integrator, and the operating frequency of the CIC filter is increased accordingly. The second method is to transform the serial structure into a parallel structure that supports multi-lane data stream processing, the processing ability of the CIC filter can be improved significantly in this way. Garcia et al. \cite{CIC_Pipeline_Class} utilizing the Residual Number System to accelerate the addition operation in the integrator part, Split operations on large bit-width data into multiple operations on small bit-width data, eventually obtain an improved speed advantage by approximately $54\%$. Teymourzadeh et al. \cite{CIC_Pipeline} proposed a pipeline CIC structure, in which the modified carry look-ahead adder and buffers between stages are exploited to speed the operation, and the final throughput of the pipeline CIC filter is up to $190$ MHz. In \cite{CIC_Heuristic}, parallel prefix adders are exploited to reduce the delay of the integrator, subsequently speeding the filtering operation. These methods mentioned above consume fewer resources, whereas the improvement in processing efficiency is limited. 
In \cite{CIC_Parallel}, a parallel CIC structure is proposed, in which a single-lane data stream is distributed to multiple CIC filters individually. This structure is designed to implement a fractional decimation factor, and the speed of the CIC filter is not improved. In \cite{CIC_Polyphase}, a polyphase CIC structure is proposed, the transfer function of the CIC filter is decomposed to the sum of multiple factors, and each factor serves as a single phase in the polyphase structure. In this way, the recursive feedback structure in the CIC filter is transformed into a parallel forward structure. Liu et al. \cite{CIC_Polyphase_SOPOT} redesigned the polyphase CIC structure to support multi-lane input, the speed of the CIC filter has been improved several times. These methods significantly improve the processing rate of the CIC filter at the cost of massive resource consumption, including multipliers, registers, and Look Up Table.

To improve the speed of the CIC filter without substantially increasing resource consumption, a novel parallel CIC structure is proposed in this paper. This method has two benefits compared with the parallel CIC structure mentioned above. Firstly, the CIC filter is transformed to parallel form at the implementation level without changing the transfer function, hence the anti-aliasing performance hasn't been degraded. Secondly, multiplication operation is not involved in this scheme, which is important to the resource-sensitive application.
The basic idea of this algorithm is to split the integrator operation into two levels. At the first level, integral is operated on whole lanes at a single clock cycle. A matrix adder is designed to sum the data in each lane Iteratively, the number of columns in the matrix adder is equal to the lane number of input data, the number of rows is calculated as $\log_2^{L}$, where \emph{L} is the lane number. At the second level, the integral is operated on each lane at the adjacent clock cycle. A column of adders and a buffer register is utilized to sum each lane, the buffer register is used to store the integral result at the previous clock cycle, the adder is used to sum the buffer data with the integral result from the first level. In this way, the serial recursive structure of CIC filter is transformed to parallel-pipeline recursive structure, integral operation within multi-lane data is realized.

Considering the wide transition band of the CIC filter, a cascade halfband filter was added after the CIC filter to improve the overall frequency response of the SRC module. The halfband filter was firstly introduced by Bellanger et al \cite{Halfband_Classical}, it can be used to decimate or interpolate digital signal by a factor of $2$. A halfband filter is a special finite impulse response(FIR) filter with a steep transition band, the corresponding coefficient is symmetry, and half of the coefficient is zero, therefore the multiplier resource required to construct the halfband filter is half of the normal FIR filter counterpart. Unfortunately, the halfband filter doesn't directly support the process of the parallel data stream. Crochiere et al. proposed a polyphase implementation structure for the FIR filter to support parallel data processing \cite{Polyphase_Classical}, this structure exploits the symmetry of the coefficient of the FIR filter, reducing the resource consumption of multipliers by half. To further utilize the zero coefficient of the halfband filter,a two path halfband filter structure was designed. The halfband filter is divided into two paths: an odd path and an even path, corresponding to the coefficients in odd and even positions respectively. The odd part is implemented in polyphase form. The even part contains only a valid coefficient of $0.5$ in the central position, so it can be implemented by a single shift register. Finally, the sum of the odd part and the even part construct the final result of the halfband filter.
\begin{figure*}[!t]
	\centering
	\includegraphics{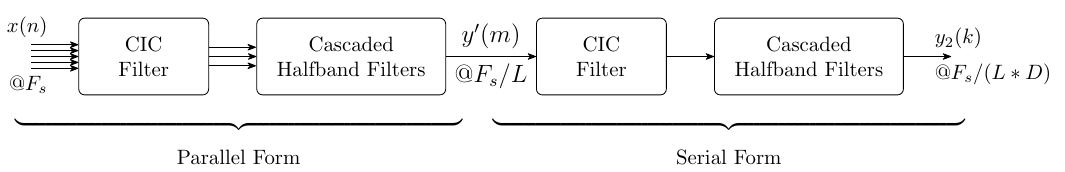}
	\caption{Structure of the Proposed cascaded Parallel-serial SRC structure. It is composed of a parallel form SRC module and a serial form SRC module, both parallel and serial parts are constructed by a CIC filter and cascaded halfband filter. The CIC filter realizes high rate decimation, and the cascaded halfband filter is used to optimize the transition band characteristics of the designed SRC filter.}
	\label{fig PSRC}
\end{figure*}

Based on the proposed parallel structures of CIC and halfband filter, a cascaded parallel-serial SRC structure was designed as shown in Fig. \ref{fig PSRC}. This parallel-serial SRC structure is composed of a parallel-form SRC module and a serial-form SRC module. The parallel part plays the role to process the parallel data stream and output the decimated single-lane data stream, the serial part decimates the single-lane data at a more flexible decimation factor. Both parallel and serial part are constructed by a CIC filter and cascaded halfband filters. The CIC filter realizes a high rate decimation, and the cascaded halfband filters is used to optimize the transition band characteristics of the designed SRC filter. Specially, the main contributions of this work are summarized as follows,
\begin{itemize}
	\item The serial recursive structure of the CIC filter is transformed to a parallel-pipeline recursive structure, which significantly improves the processing rate of the CIC filter. 
	\item A polyphase halfband filter structure is designed, which reduces the resource consumption and the design complexity of a parallel halfband filter.
	\item A parallel-serial SRC structure is proposed, which combines the advantages of high throughput of the parallel SRC and high flexibility of the serial SRC.

\end{itemize}

The rest of this paper has been organized as follows: the proposed SRC structure is described in Section II. A design example and the simulation results are given in Section III. The experimental results are presented in Section IV. Finally, conclusions are drawn in Section V.
\section{Proposed SRC Structure}
In this section, an overview of the proposed parallel-serial SRC structure is firstly presented, and several challenges in design and implementation are introduced. Then, each submodule in this SRC structure is elaborated individually, including parallel CIC, parallel halfband, and cascaded serial CIC-halfband filters.
\subsection{Architecture of the proposed SRC}

\begin{figure}[!t]
		\centering
		\includegraphics{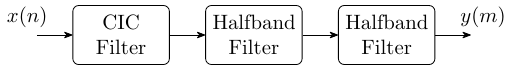}
		\caption{The block diagram of a typical SRC structure, including a CIC filter with two cascaded halfband filters.}
		\label{fig SRC}
	\end{figure}

The block diagram of a typical SRC module is shown in Fig. \ref{fig SRC}, it is composed of a CIC decimation filter and a two-stage halfband decimation filter \cite{CIC_Halfband}. The CIC decimation filter is used to realize high ratio decimation, it has the advantages of multiplierless realization and high stopband attenuation, but has a slow transition band that influences the anti-alias performance. The halfband decimation filter has a fixed decimation factor of two, it consumes few multiplier resources but has a steep transition band. Consequently, the halfband filter is always used together with the CIC filter to optimize the overall transition band response.

Although the CIC-halfband structure has the above advantages in terms of resource consumption and frequency response, its computation rate is limited and it does not directly support the processing of high-speed parallel data streams. The speed limit of the CIC filter is mainly due to its serial recursive loop structure, while the speed limit of the half-band filter is mainly due to multiplier resource consumption. In this study, the CIC serial recursive feedback structure is transformed into a parallel recursive structure, and the improved parallel CIC structure fully supports parallel data processing. At the same time, the symmetry coefficient and zero coefficient characteristics of the half-band filter is used to design a two-path polyphase halfband structure, so that the halfband filter can support parallel data processing, and the resource consumption has been reduced by 75\%.

To achieve the flexible adjustment of the decimation rate, a serial cascade CIC-halfband structure is implemented, which can realize the flexible adjustment of the decimation rate between $1$ and $32000$. Based on the parallel CIC, parallel halfband, and serial  CIC-halfband structure mentioned above, the parallel-serial SRC structure is designed and implemented as shown in Fig. \ref{fig PSRC}. In the following subsections, the parallel CIC, parallel halfband, and serial CIC-halfband structures are discussed individually.

\subsection{Architecture of the proposed Parallel CIC filter}
The structure of the CIC decimation filter is shown in Fig. \ref{fig CIC}, which is composed of three basic units: integrator, downsampler, and comb. The CIC filter has three parameters: stages number($N$), decimation ratio($R$) and differential delay($D$). $N$ corresponds to the stages of integrator and comb, $R$ is the decimation ratio of downsampler, and $D$ is the delay value of the delay unit in the comb.

The typical CIC filter structure is designed for processing serial input sequence, the speed is limited by the rate of the single lane input. One way to improve the processing rate of the CIC filter is to support the parallel data processing, and the three basic units of CIC filter should all be transformed into parallel form. In the following, the parallel integrator, parallel downsampler and parallel comb are introduced individually.
\subsubsection{Parallel Integrator}
\begin{figure}[!t]
		\centering
		\includegraphics{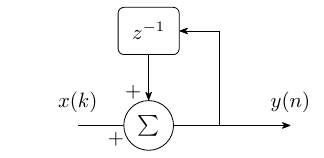}
		\caption{The structure of integrator unit, it is composed of a delay unit and an accumulator.}
		\label{fig CICInt}
	\end{figure}

The integrator unit is composed of a delay unit and an accumulator as shown in Fig. \ref{fig CICInt}. 
It has a serial recursive structure, and cannot directly support parallel data processing. In the following, this serial recursive structure is transformed into parallel recursive structure. 

\begin{figure}[!t]
	\centering
	\includegraphics{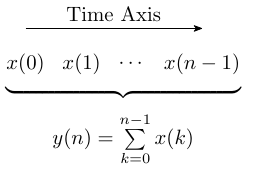}
	\caption{Integral operation on serial sequence along the time axis, accumulate all the data from the beginning.}
	\label{fig SInt_Map}
\end{figure}

The operation of integrator can be expressed as 
\begin{equation} 
	y(n) = \sum_{k=0}^{n-1}{x(k)}
	\label{eq 0}
\end{equation}
where $k$ is the time index, $ x(k) $ and $ y(n) $ are the input sequence and integral result, respectively. In this situation, the integrator operation is executed along the time axis, the integral process is shown in Fig. \ref{fig SInt_Map}.

\begin{figure}[!t]
	\centering
	\includegraphics{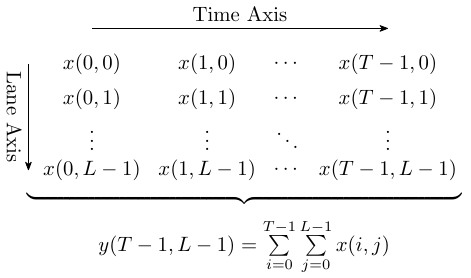}
	\caption{Integral operation on parallel sequence along both time axis and lane axis, the parallel sequence is mapped from serial sequence directly.}
	\label{fig PInt_Map}
\end{figure}

To obtain the parallel form of integrator, the serial sequences $x(k)$ and $y(n)$ can be mapped to parallel sequences $x(i,j)$ and $y(i,j)$ individually.
\begin{IEEEeqnarray}{rCl}
	x(k) & \xrightarrow[j=k\bmod L]{i=\lfloor \cfrac{k}{L} \rfloor} & x(i,j)\\
	y(n) & \xrightarrow[j=n\bmod L]{i=\lfloor \cfrac{n}{L} \rfloor} & y(i,j)
\end{IEEEeqnarray}
where $i$ is the time index, $i=0,1,...,T$, and $T$ is the current time index.  $j$ is the lane index, $j=0,1,...,L-1$, and $L$ is the lane number of parallel sequence. 
The mapped parallel sequence is shown in Fig. \ref{fig PInt_Map}. The horizontal axis of parallel sequence is the time $i$. The vertical axis is the lane $j$. The input sequence $\bm{x(t)}$ and output sequence $\bm{y(t)}$ at time $t$ can be represented as
\begin{equation}
	\bm{x(t)} = 
	\begin{pmatrix}
		x(t,0)\\
		x(t,1)\\
		\vdots\\
		x(t,L-1)\\
	\end{pmatrix}\hspace{5mm}
	\bm{y(t)} = 
	\begin{pmatrix}
		y(t,0)\\
		y(t,1)\\
		\vdots\\
		y(t,L-1)\\
	\end{pmatrix}
	\label{Eq PInt IOVector}
\end{equation} 
The integral operation is performed on both time axis and lane axis. In the following, the integral operation on time and lane axis is called time integral and lane integral respectively.

The lane integral at time $ t $ is denoted as $ I(t,l) $. So
\begin{equation}
	I(t,l) = \sum\limits_{j=0}^{l}{x(t,j)}
\end{equation}
Then the integral result $y(t,l)$ at time $t$ can be derived as
\begin{IEEEeqnarray}{rCl}
		y(t, l) &=&y(L*t+l) \nonumber\\ 
		&=&\sum\limits_{k=0}^{L*t+l}{x(k)} \nonumber\\
		&=&\sum\limits_{i=0}^{t-1}{\sum\limits_{j=0}^{L-1}{x(L*i+j)}}+\sum\limits_{j=0}^{l}{x(L*t+j)} \nonumber\\
		&=&\sum\limits_{i=0}^{t-1}{\sum\limits_{j=0}^{L-1}{x(i,j)}}+\sum\limits_{j=0}^{l}{x(t,j)} \nonumber\\
		&=&\sum\limits_{i=0}^{t-1}{I(i,L-1)} + I(t,l)
	\label{Eq PInt}
\end{IEEEeqnarray}
where $l=0,1,...,L-1$. 
substituting (\ref{Eq PInt}) into (\ref{Eq PInt IOVector}), the output of parallel integrator can be represented as
\begin{equation}
	\bm{y(t)} = 
	\begin{pmatrix}
		y(t,0)\\
		y(t,1)\\
		\vdots\\
		y(t,L-1)\\
	\end{pmatrix}=
	\sum\limits_{i=0}^{t-1}{I(i,L-1)}+
	\begin{pmatrix}
		 I(t,0)\\
		I(t,1)\\
		\vdots\\
		I(t,L-1)\\
	\end{pmatrix}
	\label{Eq PInt InOut}
\end{equation} 

\begin{figure}[!t]
	\centering
	\includegraphics{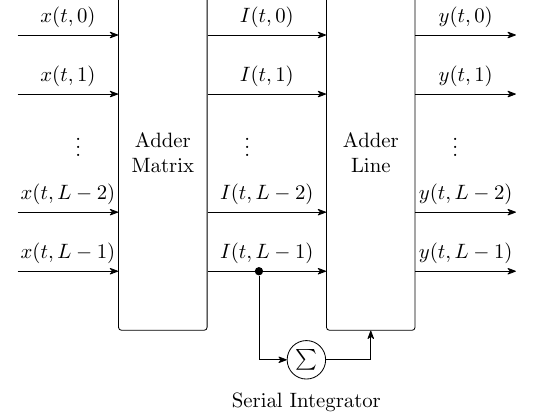}
	\caption{The proposed parallel CIC structure includes an adder matrix and an adder line. The adder matrix calculates the integral along the lane axis, and the adder line calculates the integral along the time axis.}
	\label{fig PCIC}
\end{figure}

In other words, the calculation of $\bm{y(t)}$ can be divided into three steps. The integral of $ I(i,L-1) $ along time axis, the calculation of $ I(t,l) $, and finally the summation of above two. The resulting parallel integral structure is derived as Fig. \ref{fig PCIC}.
In this parallel integral structure, the adder matrix is used to calculate the $ I(t,l) $ , the serial integrator is used to calculate the integral result of $ I(i, L-1) $ along the time axis, and the adder line adds up these two-part. In the following, these three parts are described individually.
 
The adder matrix is used to calculates the result of $I(t,l)$, where $l=0,1,...,L-1$. It can be described as 
\begin{equation}
	\bm{I(t)} = 
	\begin{pmatrix}
		I(t,0)\\
		I(t,1)\\
		\vdots\\
		I(t,L-1)\\
	\end{pmatrix}=
	\begin{pmatrix}
		\sum\limits_{j=0}^{0}{x(t,j)}\\
		\sum\limits_{j=0}^{1}{x(t,j)}\\
		\vdots\\
		\sum\limits_{j=0}^{L-1}{x(t,j)}\\
	\end{pmatrix}
	\label{Eq PInt Integral}
\end{equation}
where $\bm{I(t)}$ is the output vector of the adder matrix at time $t$.
For the case that $L$ is even, $\bm{I(t)}$ can be divided into two short vector
\begin{equation}
	\bm{I_a(t)} = 
	\begin{pmatrix}
		I(t,0)\\
		I(t,1)\\
		\vdots\\
		I(t,L/2-1)\\
	\end{pmatrix}=
	\begin{pmatrix}
		\sum\limits_{j=0}^{0}{x(t,j)}\\
		\sum\limits_{j=0}^{1}{x(t,j)}\\
		\vdots\\
		\sum\limits_{j=0}^{L/2-1}{x(t,j)}\\
	\end{pmatrix}
	\label{Eq Int Top}
\end{equation}
\begin{equation}
	\bm{I_b(t)} = 
	\begin{pmatrix}
		I(t,L/2)\\
		I(t,L/2+1)\\
		\vdots\\
		I(t,L-1)\\
	\end{pmatrix}=
	\begin{pmatrix}
		\sum\limits_{j=0}^{L/2}{x(t,j)}\\
		\sum\limits_{j=0}^{L/2+1}{x(t,j)}\\
		\vdots\\
		\sum\limits_{j=0}^{L-1}{x(t,j)}\\
	\end{pmatrix}
	\label{Eq Int Bot}
\end{equation}
where $\bm{I_a(t)}$ is the top half of the $\bm{I(t)}$, and $\bm{I_b(t)}$ is the bottom half. $\bm{I_a(t)}$ is actually the lane integral operation on the top half of $\bm{x(t)}$. And the common elements $\sum\limits_{j=0}^{L/2-1}{x(t,j)}$ of the vector $\bm{I_b(t)}$ can be extracted
\begin{equation}
	\bm{I_b(t)} = 
	\sum\limits_{j=0}^{L/2-1}{x(t,j)}+
	\begin{pmatrix}
		\sum\limits_{j=L/2}^{L/2}{x(t,j)}\\
		\sum\limits_{j=L/2}^{L/2+1}{x(t,j)}\\
		\vdots\\
		\sum\limits_{j=L/2}^{L-1}{x(t,j)}\\
	\end{pmatrix}
	\label{Eq Int BotSplit}
\end{equation}
The first part is equal to $I(t,L/2-1)$, it is actually the last element of $\bm{I_a(t)}$. the second part is the lane integral operation on the bottom half of $\bm{x(t)}$. 

For simplicity, the lane integral operation on top half of $\bm{x(t)}$ is denoted as $\bm{I_2^0(t)}$, lane integral on bottom half of $\bm{x(t)}$ is denoted as $\bm{I_2^1(t})$, the subscript $2$ means the integral operation is divided into two groups equally, the superscript is the index of the selected group. The two groups $\bm{I_2^0(t)}$ and $\bm{I_2^1(t)}$ can be represented as
\begin{equation}
	\bm{I_2^0(t)} = 
	\begin{pmatrix}
		\sum\limits_{j=0}^{0}{x(t,j)}\\
		\sum\limits_{j=0}^{1}{x(t,j)}\\
		\vdots\\
		\sum\limits_{j=0}^{L/2-1}{x(t,j)}\\
	\end{pmatrix}
	\hspace{3mm}
	\bm{I_2^1(t)} = 
	\begin{pmatrix}
		\sum\limits_{j=L/2}^{L/2}{x(t,j)}\\
		\sum\limits_{j=L/2}^{L/2+1}{x(t,j)}\\
		\vdots\\
		\sum\limits_{j=L/2}^{L/2-1}{x(t,j)}\\
	\end{pmatrix}
	\label{Eq GroupDef}
\end{equation}
Substituting (\ref{Eq GroupDef}) into (\ref{Eq PInt Integral}), (\ref{Eq Int Top}) and (\ref{Eq Int BotSplit}). $\bm{I(t)}$ can be represented as
\begin{equation}
	\bm{I(t)} = 
	\begin{pmatrix}
		\bm{I_a(t)}\\
		\bm{I_b(t)}\\
	\end{pmatrix} = I_2^0(t,L/2-1)
	\begin{pmatrix}
		\bm{0_{L/2}}\\
		\bm{1_{L/2}}\\
	\end{pmatrix} + 
	\begin{pmatrix}
		\bm{I_2^0(t)}\\
		\bm{I_2^1(t)}\\
	\end{pmatrix}
	\label{Eq GroupAdd}
\end{equation}
where $\bm{0_{L/2}}$ is a zero vector of length $L/2$, $\bm{1_{L/2}}$ is a all ones vector of length $L/2$, and $I_2^0(t,L/2-1)$ is the last element of $\bm{I_2^0(t)}$.
\begin{equation}
	I_2^0(t,L/2-1) = \sum\limits_{j=0}^{L/2-1}{x(t,j)}
\end{equation}

\begin{figure}[!t]
	\centering
	\includegraphics{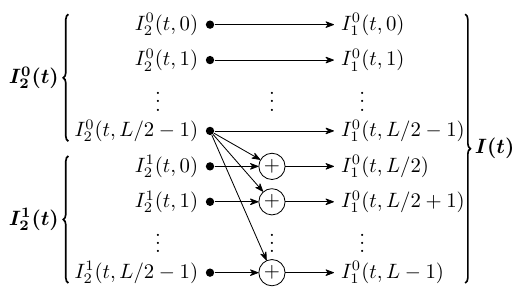}
	\caption{Decompose the long sequence integral into two short sequence integrals.}
	\label{Fig AddMatrixUnits}
\end{figure}

That is to say, the integral operation on a sequence of even length can be divided into the integral operation on two shorter sequence, the corresponding implementation structure is shown in Fig. \ref{Fig AddMatrixUnits}, where “$+$” is an add operation.

\begin{figure*}[!t]
	\centering
	\includegraphics{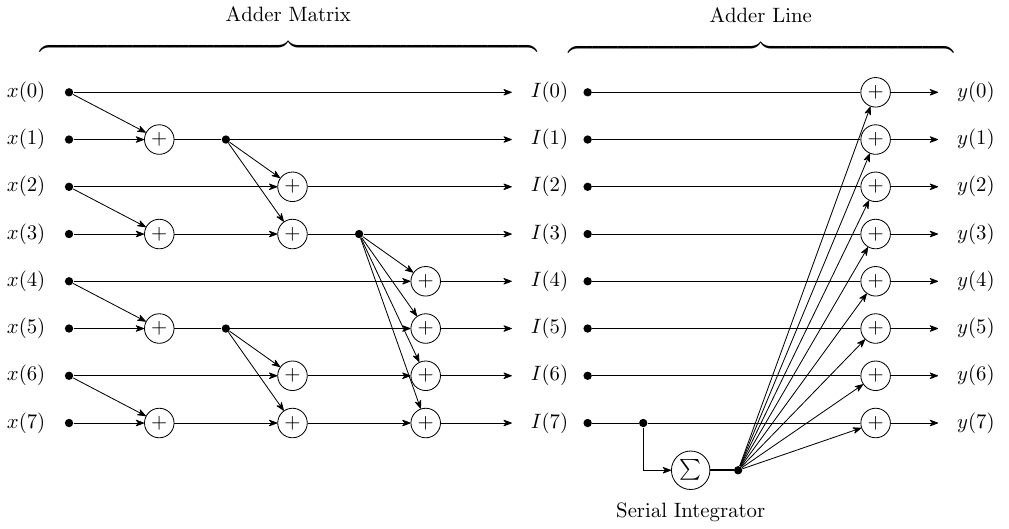}
	\caption{The implementation structure of the adder matrix and adder line for an 8-lane parallel integrator. This structure can be expanded to any line integrator.}
	\label{Fig PInt}
\end{figure*}

 An example of 8 lane adder matrix is shown in Fig. \ref{Fig PInt}, where the left part is the adder matrix.

For the case that $L$ is not a power of 2, two step can be adopted to obtain the result. In the first step, an adder matrix of $\lceil log2^{(L)} \rceil$ can be derived. the output of the adder matrix is 
\begin{equation}
	\bm{I_1^0} = 
	\begin{pmatrix}
		\sum\limits_{j=0}^{0}{x(j)}\\
		\sum\limits_{j=0}^{1}{x(j)}\\
		\vdots\\
		\sum\limits_{j=0}^{\lceil log2^{(L)} \rceil-1}{x(j)}\\
	\end{pmatrix}
\end{equation}
note that the top $L$ lanes are the desired result. As shown in Fig. \ref{Fig AddMatrixUnits}, the calculation of $I_1^0{(l)}$ is irrelevant to the last $N-1-l$ lane of the left side. In other words, the top $L$ lanes of the output don't rely on the last $ \lceil log2^{(L)} \rceil - L $ lanes of the input. Therefore, in the second step the last $\lceil log2^{(L)} \rceil - L$ lanes can be removed directly.

The accumulator calculate the integral of $ I(i,L-1) $ along time axis. The input of the accumulator at time $i$ is $I(i,L-1)$, the output of accumulator at time $t$ is defined as $A(t)$, thus  
\begin{equation}
	A(t)=\sum\limits_{i=0}^{t}{I(i,L-1)}
\end{equation}
The result of $A(t)$ is exactly the first part of (\ref{Eq PInt}). Finally, the adder line adds up the output of the accumulator and the output of the adder matrix for each lane. A complete structure of the parallel integral for 8 lane sequence is shown in Fig. \ref{Fig PInt}.

\subsubsection{Parallel Downsampler}
\subsubsection{Parallel Comb}
\begin{figure}[!t]
	\centering
	\includegraphics{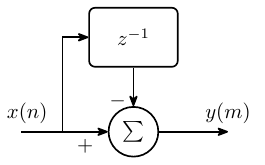}
	\caption{The comb in the CIC filter includes an adder and a delay unit.}
	\label{fig CIC_Comb}
\end{figure}

The comb part of the CIC filter should also be transformed into a parallel form to support parallel data processing. The structure of the serial comb part is shown in Fig. \ref{fig CIC_Comb}, it includes a subtractor and a delay unit.
The input parallel sequence is denoted as $ x(t,l) $, the parallel output $ y(t,l) $ can be derived as follow.
\begin{equation}
	\begin{array}{lll}
		y(t,l)&=& y_1(t*L+l)\\
		 &=& x_1(t*L+l) - x_1(t*L+l-M)\\
		 &=& x(t,l) \\
		 && -x(\lfloor(t*L+l-M)/L\rfloor,(t*L+l-M)\%L)
	\end{array}
	\label{equa CombM}
\end{equation}

where $M$ is the differential delay of the CIC filter. Considering that the parameter $M$ is typically set to $1$, the parallel output of each lane can be simplified as follow.
\begin{equation}
	\begin{array}{ll}
		\begin{pmatrix}
			y(t,0)\\
			y(t,1)\\
			\vdots \\
			y(t,L-2)\\
			y(t,L-1)\\
		\end{pmatrix}
		=
		\begin{pmatrix}
			x(t,0) - x(t-1,L-1)\\
			x(t,1) - x(t,0)\\
			\vdots \\
			x(t,L-2) - x(t,L-2)\\
			x(t,L-1) - x(t,L-1)\\
		\end{pmatrix}
	\end{array}
	\label{eqau Com1}
\end{equation}

\begin{figure}[!t]
	\centering
	\includegraphics{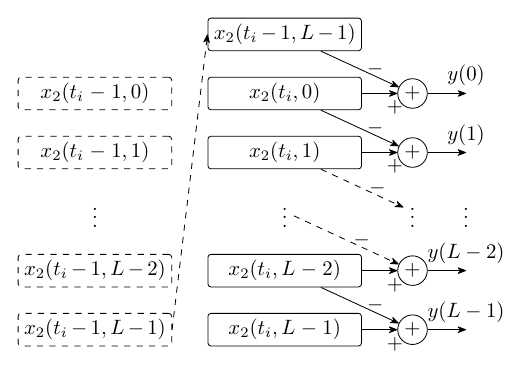}
	\caption{In the structure of the proposed parallel comb, multiple subtractor operations are conducted simultaneously.}
	\label{fig PComb}
\end{figure}

The final parallel comb structure derived from (\ref{eqau Com1}) is shown in Fig. \ref{fig PComb}.

\subsection{Architecture of the proposed parallel halfband filter}
In this subsection, the parallel structure of an FIR filter is firstly derived, and then the property of halfband filter is utilized to reduce the resources consumption.
\subsubsection{Parallel Structure for FIR Filter}
\begin{figure}[!t]
	\centering
	\includegraphics{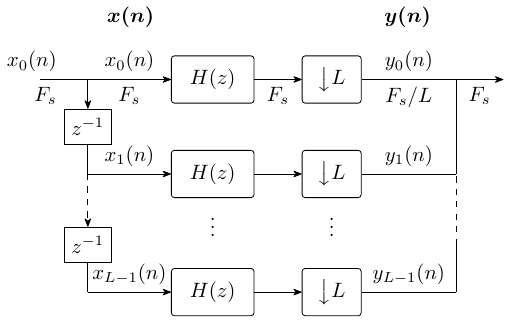}
	\caption{The implementation structure of the parallel FIR filter, the input sequences are distributed to L lane after time shifting.}
	\label{Fig PFIR}
\end{figure}

An FIR filter of length N with input $ x(n) $ and output y(n) can be described as follow.
\begin{equation}
	y(n) = \sum\limits_{k=0}^{N-1}{h(k)\cdot x(n-k)} = h(n)\otimes x(n)
\end{equation}
the $z$ transform of $y(n)$ is
\begin{equation}
	Y(z) = z\{y(n)\}=z\{h(n)\otimes x(n)\} = H(z)X(z)
\end{equation}
To obtain the parallel form of this filter, the $x(n)$ and $y(n)$ can be mapped to a parallel form.
\begin{equation}
	\begin{array}{l}
		y_j(n) \rightarrow y(n*L-j)\\
		x_j(n) \rightarrow x(n*L-j)\\
	\end{array}
\end{equation}
where $j=0,1,...L-1$. $y_j(n)$ at $j$th lane can be derived as
\begin{equation}
	\begin{array}{ll}
		y_j(n) &= y(n*L-j)\\
		&=\sum\limits_{k=0}^{N-1}{h(k)x(n*L-j-k)}\\
		&=\sum\limits_{k=0}^{N-1}{h(k)x_j(n-k)}
	\end{array}
\end{equation}
Thus the z transform of $y_j(n)$ can be derived as
\begin{equation}
	\begin{array}{ll}
		Y_j(z) 
		&=z\{y(n*L-j)\}\\ 
		&=z^{-j}Y(z^{1/L})\\
		&=z^{-j}H(z^{1/L})X(z^{1/L})	
	\end{array}
	\label{Eq ZTrans}
\end{equation}
The parallel form of $\bm{x(n)}$ and $\bm{y(n)}$ can be represented as (\ref{Eq PFIRMap}).
\begin{equation}
	\begin{array}{ll}
		\bm{x(n)} = 
		\begin{pmatrix}
			x_0(n)\\
			x_1(n)\\
			\vdots\\
			x_{L-1}(n)\\
		\end{pmatrix}\ \ 
		\bm{y(n)}=
		\begin{pmatrix}
			y_0(n)\\
			y_1(n)\\
			\vdots\\
			y_{L-1}(n)\\
		\end{pmatrix}
	\end{array}
	\label{Eq PFIRMap}
\end{equation}
Substituting (\ref{Eq ZTrans}) into (\ref{Eq PFIRMap}), the z transform of $\bm{y(n)}$ can be obtained as (\ref{Eq PFIR}).
\begin{figure*}[!t]
\begin{equation}
	\begin{array}{l}
		\bm{Y(z)}=z\{\bm{y(n)}\}
		=
		\begin{pmatrix}
			z\{y_0(n)\}\\
			z\{y_1(n)\}\\
			\cdots\\
			z\{y_{L-1}(n)\}
		\end{pmatrix}
		=\begin{pmatrix}
			Y(z^{1/L})\\
			z^{-1}Y(z^{1/L})\\
			\vdots\\
			z^{-L+1}Y(z^{1/L})\\
		\end{pmatrix}
		=\begin{pmatrix}
			H(z^{1/L})X(z^{1/L})\\
			z^{-1}H(z^{1/L})X(z^{1/L})\\
			\vdots\\
			z^{-L+1}H(z^{1/L})X(z^{1/L})\\
		\end{pmatrix}
	\end{array}
	\label{Eq PFIR}
\end{equation}
\end{figure*}
Eventually, the parallel FIR structure derived from (\ref{Eq PFIR}) is shown in Fig. \ref{Fig PFIR}.
\subsubsection{Parallel Structure for Halfband Filter}
\begin{figure}[!t]
		\centering
		\includegraphics[width=2.5in]{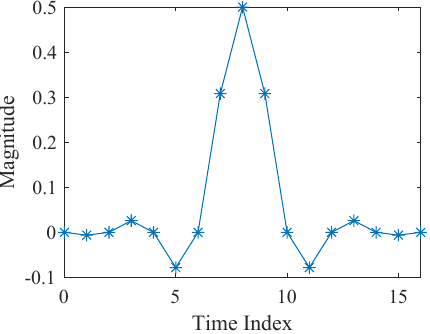}
		\caption{The impulse response of a halfband filter, which is symmetry about the center point, has zero values at even index offsets.}
		\label{Response_HB}
\end{figure}

In the second step, the coefficients property of halfband filter was applied to the parallel structure.
The impulse response of a lowpass halfband filter is.
\begin{equation}
	h_{LP}=\cfrac {1}{2}\cfrac{sin(\cfrac {n \pi }{2})}{\cfrac {n \pi }{2}}
	\label{Impuse_HB}
\end{equation}
It contains one sample at the point of symmetry with matching left and right samples about the symmetry point, as shown in Fig. \ref{Response_HB}. The impulse response samples of the halfband filter are seen to be zero at the even index offsets from the center point of the filter, and the samples with odd index offset are seen to exhibit even symmetry about the filter center point.

The transfer function corresponds to a halfband filter  of length $2N+1$ is given by
\begin{equation}
	H(z) = \sum\limits_{k=0}^{2N}{h(k)z^{-k}}
\end{equation}
and the following equation is satisfied.
\begin{equation}
	h(k) = 
	\begin{cases}
		h(2N-k), & {\text{if}}\ (k-N)\%2=1\\
		0.5, & {\text{if}}\ k=N\\
		0, & {\text{otherwise.}}
	\end{cases}
	\label{HB_Property}
\end{equation}
In (\ref{HB_Property}), 
When the $N$ is even, $(k-N)\%2=k\%2$, when $N$ is odd, $(k-N)\%2=(k+1)\%2$. In other words, for even $N$, coefficients at even index except $N$ are zeros, coefficients at odd index are symmetry, for odd $N$, coefficients at odd index except $N$ are zeros, coefficients at even index are symmetry.
These properties permit the non-polyphase form of the filter of length $2N+1$ to be implemented with only $N/2$ multiplications per output sample \cite{CIC_Halfband}.
In the following, the zero and symmetry coefficients is firstly applied to a serial structure, and then the serial structure is transformed into a parallel structure.

\begin{figure}[!t]
		\centering
		\includegraphics{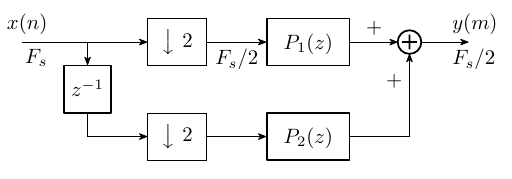}
		\caption{In the structure of the proposed two-path parallel halfband filter, the first lane is the delay part, and the lanes below can be treated as a polyphase FIR filter.}
		\label{fig PHB}
\end{figure}

By utilizing the zero coefficient of the halfband filter, a two-path structure can reduce the multiplier resource consumption of the halfband filter by $50\%$.
An FIR filter of length $2N+1$ with input $ x(n) $ and output $y(n)$ can be described as follow.
\begin{equation}
		y(n) = \sum\limits_{k=0}^{2N}{h(k)\cdot x(n-k)}
	\label{HB_IO}
\end{equation}
where the $h(k)$ are filter coefficients. Firstly the case of even N is considered, and $ y(n) $ can be represented as follow.
\begin{equation}
	\begin{array}{lll}
	y(n)&=&\sum\limits_{\substack{k=0\\k\ne N/2}}^{N}{h(2k)\cdot x(n-2k)} + h(N)\cdot x(n-N)\\
	&&+\sum\limits_{k=0}^{N-1}{h(2k+1)\cdot x(n-2k-1)}\\
	&=&\cfrac {x(n-N)}{2}+\sum\limits_{k=0}^{N-1}{[h(2k+1) \cdot x(n-2k-1)]}  \\
	&=&P_1 + P_2\\
	\end{array}
\end{equation}
Where $ P_1 $ can be implemented by a shift register and delay unit, $ P_2 $ is equivalent to an FIR filter of length $N$ and can be implemented in a polyphase structure. The two-path parallel structure composed by $P_1$ and $ P_2 $ is shown in Fig. \ref{fig PHB}.

\begin{figure}[!t]
	\centering
	\includegraphics{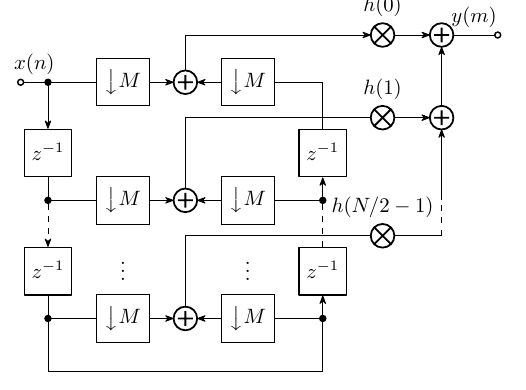}
	\caption{The symmetry polyphase structure utilizing the symmetry property of the coefficients, half of the multipliers can be saved in this structure. Data at symmetry index offset are added firstly, and then multiplied with the common coefficient.}
	\label{Polyphase_Symmetry}
\end{figure}

Secondly, to exploit the symmetric coefficients of the halfband filter, the resource consumption can be further reduced by 50\%. For simplicity, $x(n-2k-1) $ is defined as $x'(2k+1)$, $x$ and $x'$ exhibit even symmetry about point of $n/2$. For even values of N, $ E_1 $ can be represented as follow.
\begin{equation}
	\begin{array}{lll}
		E_1&=&\sum\limits_{k=0}^{N/2-1}{[h(2k+1) \cdot x'(2k+1)]}\\
		&&+\sum\limits_{k=N/2}^{N-1}{[h(2k+1) \cdot x'(2k+1)]}\\
	\end{array}
	\label{Eq E1}
\end{equation}
Note that $ h(2k+1) = h(2N-2k-1) = h(2(N-1-k)+1) $, substitute $ k=N-1-t $ in (\ref{Eq E1}).
\begin{equation}
	\begin{array}{ll}
		E_1&=\sum\limits_{k=0}^{N/2-1}{h(2k+1) \cdot [x'(2k+1) + x'(2N-2k-1)]}\\
	\end{array}
\end{equation}
Thus, $E_1$ can be implemented in a symmetric polyphase structure as shown in Fig. \ref{Polyphase_Symmetry}. 
Total $N/2$ multiply operations and $N/2$ add operations are involved in this structure.
	
Considering that the Xilinx IP Core supports a symmetric polyphase structure,  the IP Core is used directly to implement the operation of $ E_1 $, which further reduces the design complexity.

For odd values of $N$, the corresponding two-path parallel polyphase structure can be derived in the same way.
\subsection{Architecture of the Cascaded Serial SRC}
The serial SRC part in the parallel-serial SRC structure is used to realize flexible ratio adjustment, it is constructed by a CIC filter with two-stage halfband filters. The decimation factor of the parallel-serial SRC can be adjusted efficiently by simply changing the parameter of the serial CIC filter, and the cascaded halfband filters are used to optimize the transition band of the SRC.  In this work, the CIC filter and halfband filters are implemented by the official IP (intellectual property) of AMD-Xilinx, the corresponding IPs are CIC Compiler and FIR Compiler respectively \cite{CIC_Compiler,Halfband_Compiler}. these IPs provide flexible parameter adjustment, lower resource consumption and design complexity.
\section{Design Example with Simulation Result}
In this section, a design example of the proposed parallel-serial SRC structure is presented. Firstly, the design parameters of each filter are elaborated. And then, several simulations are conducted to evaluate the performance of the SRC.
\subsection{Design Specification}
In this design, The structure in Fig.\ref{fig PSRC} is applied, the specification of the SRC is shown in Table \ref{SRC Param}.
\begin{table}[!t]
	\centering
	\caption{Specification of the SRC}
	\begin{tabular}{lc}
		\hline  \hline	
		Characteristic  	& Description\\ \hline
		Sampling Rate  		& $20$ GSPS\\ 
		Input Lanes			&$80$\\
		Bandwidth 			& $8$ GHz \\
		Frequency Range  	&$10$ kHz to $8$ GHz  \\ 
		Decimation Ratio  	& $80$ to $2560000$ \\
		Anti-Alias  		&$\ge 60$ dB \\ 
		\hline	\hline	
	\end{tabular}
	\label{SRC Param}
\end{table}

The input signal has a fixed sampling rate of $20$ GSPS, and the lane number is set to $80$. That is to say, the sampling rate of each lane is $250$ MSPS. The input frequency range is limited to $10$ kHz to $8$ GHz. In addition, this design provides a flexible adjustment on decimation ratio ranging from $1$ to $2560000$, and the suppression on aliasing components at each decimation ratio is up to $60$ dB.
\begin{table}[!t]
	\centering
	\caption{Design Parameters of the CIC Filter}
	\begin{tabular}{lcc}
		\hline \hline
		Parameters  			& Parallel 		& Serial \\ \hline
		Decimation ratio (R)  	& $20$ 			& $1$ to $4000$\\ 
		Number of stage (N)  	& $5$ 			& $5$ \\ 
		Differential delay (M)  & $1$  			& $1$\\ 
		Input bit width			& $16$			& $16$\\
		Output bit width		& $16$			& $16$\\ 
		Rounding method			& Round			& Round\\
		\hline \hline
	\end{tabular}
	\label{CIC Param}	
\end{table}

The design parameters of the CIC filter are given in Table \ref{CIC Param}. The parallel CIC and serial CIC have the same parameter except for the decimation ratio, the parallel CIC has a fixed ratio of $20$ to reduce the design complexity, and the serial CIC has an adjustable ratio range from $1$ to $4000$. The number of stages of both CIC filters are set to $5$ to provide a $60$ dB stopband attenuation.

\begin{table}[!t]
	\centering
	\caption{Design Parameters of the halfband Filter}
	\begin{tabular}{lcc}
		\hline \hline
		Parameters				& Parallel Halfband				& Serial Halfband \\ \hline
		Cascaded stages  		& $2$ 					& $3$\\ 
		Filter rrder  			& $122$ 					& $238$\\ 
		Transition width  		& $0.03$ 				& $0.015$ \\ 
		Stop attenuation		& $70$  				& $70$\\ 
		Coefficients bit width	& $16$					& $16$\\				
		Output bit width		& $16$					& $16$\\
		Rounding Method			& $Round$				& $Round$\\ 
		\hline \hline
	\end{tabular}
	\label{HB Param}
\end{table}
The design parameters of the halfband filter are given in Table \ref{HB Param}. the stages of the parallel halfband filter are set to $2$, thus the total decimation ratio of the parallel SRC is $ 80 $. As a result, the parallel SRC output a serial data stream to the serial SRC. In addition, The order of the parallel halfband filter is set to $122$ to provide a $60$ dB stopband attenuation and $0.03$ dB transition width. Considering that the serial part has lower design complexity, higher performance can be achieved in the serial part with high stages and filter order.
\subsection{Simulation Results}
Several simulations were conducted to assess the anti-aliasing performance of the SRC,  the magnitude response and the system response of the SRC are presented in the following.

\begin{figure}[!t]
		\centering
		\includegraphics[width=3.5in]{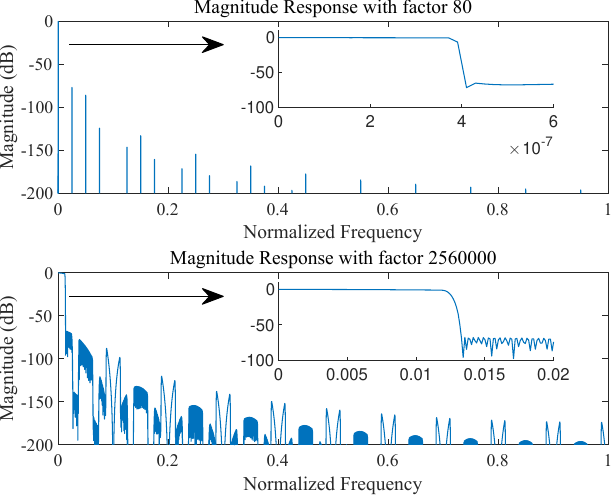}
		\caption{Magnitude Response with factor 80 and 2560000.}
		\label{Response_SRC}
\end{figure}

Firstly, the amplitude response of the SRC at each decimation ratio has been analyzed. Considering that more than $4000$ decimation ratios are supported in this design, only the first five ratios and the last five ratios are listed in Table \ref{Perf_Decim}. all the passband ripple is lower than $0.2dB$, and the stopband attenuation is higher than $70dB$. The amplitude response of the SRC with factors $80$ and $2560000$ are shown in Fig. \ref{Response_SRC}.
\begin{table}[!t]
	\centering
	\caption{Design Parameters for the typical decimation factor}
	\begin{tabular}{lcc}
		\hline \hline
		Decimation Ratio		& Passband Ripple	& Stopband Attenuation \\ \hline
		$80$  					& $0.7241$ dB 		& $67.8443$ dB\\ 
		$160$  					& $0.1876$ dB 		& $65.1379$ dB\\ 
		$320$  					& $0.0667$ dB 		& $64.8181$ dB \\ 
		$640$					& $0.0443$ dB  		& $64.73922$ dB\\ 
		$1600$					& $0.7152$ dB		& $66.3513$ dB\\ 
		$3200$					& $0.1993$ dB		& $65.1525$ dB\\ 
		$3840$					& $0.1982$ dB		& $65.1555$ dB\\ 
		$4480$					& $0.1982$ dB		& $65.1701$ dB\\ 
		$5120$					& $0.1980$ dB		& $65.1570$ dB\\ 
		\vdots					& \vdots			& \vdots\\
		$\ge 5760$				& $\le 0.2000$ dB	& $\ge 65$ dB\\ 
		\hline \hline
	\end{tabular}
	\label{Perf_Decim}
\end{table}

\begin{figure}[!t]
		\centering
		\includegraphics[width=3.5in]{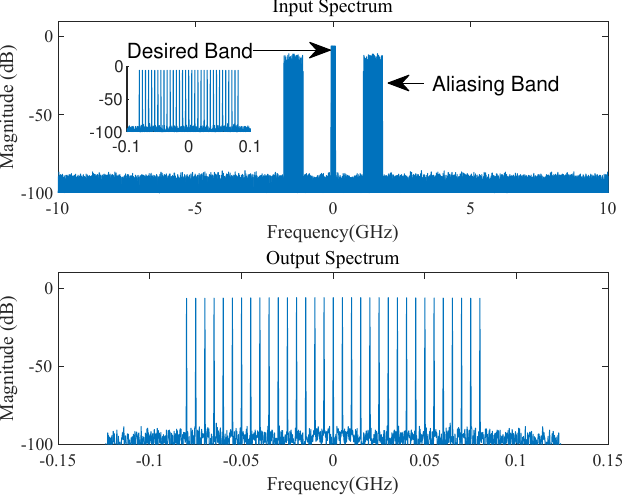}
		\caption{The simulation was conducted to verify the anti-aliasing performance of the design SRC, the aliasing band on the top plot has been attenuated by nearly 100dB as shown in the bottom plot.}
		\label{Response_Input}
\end{figure}

Secondly, the system response of the SRC has been verified to evaluate the anti-aliasing effectiveness. 
 In this example, the SRC with factor 80 is considered. The desired signal of the SRC is a multitone signal, composed of frequencies ranging from $10$ MHz to $80$ MHz in $10$ MHz intervals. The amplitude of each sinusoid component in the multitone signal is equal and normalized to $1$. There also exists an unwanted signal that will alias on top of the desired signal due to SRC. The spectrum of the input signal and decimated signal is shown in Fig. \ref{Response_Input}. 

The spectrum of the input signal is shown in the top plot, the desired band is centered at zero frequency and the aliasing band at $F_{out}$, the aliasing band will fold on top of the desired band as a result of the decimation process. The bottom plot shows the spectrum of the decimated signal, it can be seen that the desired band is intact, whereas the aliasing is the noise floor, having been attenuated by more than $70$ dB.

\section{Experimental Results and Analysis}
\begin{table*}[!t]
	\centering
	\caption{Resource Consumption of the designed SRC}
	\begin{tabular}{lcccccccccc}
		\hline \hline
		\multirow{2}*{Resources} 	& Parallel CIC filter 	& \multicolumn{2}{c}{Parallel halfband filter} 	& Serial CIC filter    	& \multicolumn{3}{c}{Serial halfband filter}	&Total	&Utilization\\ 
		&   					& First stage	& Second stage				& 			 				& First stage 	& Second stage 	& Third stage	&\\\hline 
		LUTs 							& $37735$ 				& $838$			& $669$						& $1355$					& $935$			& $935$  		& $935$			& $43402$	&$6.54\%$\\ 
		FFs  							& $59275$ 				& $3083$		& $2122$					& $1628$					& $5337$ 		& $5337$ 		& $5337$		& $82119$	&$6.19\%$\\ 
		DSP48Es							& $0$					& $62$			& $31$						& $0$						& $30$ 			& $30$ 			& $30$			& $183$	&$3.32\%$\\ 
		\hline \hline
	\end{tabular}
	\label{Resource}
\end{table*}  
\begin{table*}[!t]
	\centering
	\caption{Performance Comparison between different methods}
	\begin{tabular}{lcccc}
		\hline\hline
		&Proposed 						& Pipelined SRC 		& Polyphase SRC			& Serial SRC\\ \hline
		Input sampling rate		&$20$ GSPS						&$7.2$ GSPS				&$7.2$ GSPS				&$-$				\\
		Input signal bandwidth  &$10$ GSPS						&$3.6$ GHz				&$3.6$ GHz				&$200$ MHz 		\\ 
		Output bandwidth		&$7.8125$ kS/s to $250$ MS/s	&$1$ kS/s to $225$ MS/s	&$28.125$ MHz			&$1$ MHz			\\
		Decimation factor		&$80$ to $2560000$				&$32$ to $8388608$		&$192$ 					&$192$			\\ 
		Passband ripple			&$\le 0.8$ dB					&$\le 0.1$ dB			&$\le 0.1$ dB			&$\le 0.1$ dB	\\
		Stopband attenuation	&$\ge 65$ dB					&$\ge 80$ dB			&$\ge 80$ dB 			&$\ge 80$ dB  	\\
		Multipliers per input  	&$\approx 2$					&$\approx 6$ 			&$40$	 				&$149$			\\ 
		LUTs per input			&$\approx 542$					&$\approx 1126$			&$1267$					&$1828$			\\
		\hline\hline	
	\end{tabular}
	\label{performance}
\end{table*}
\subsection{Implementation Platform}
\begin{figure}[!t]
		\centering
		\begin{tikzpicture} 
		\node [inner sep=0pt] (brd) at (0mm,0mm) {\includegraphics[width=2in]{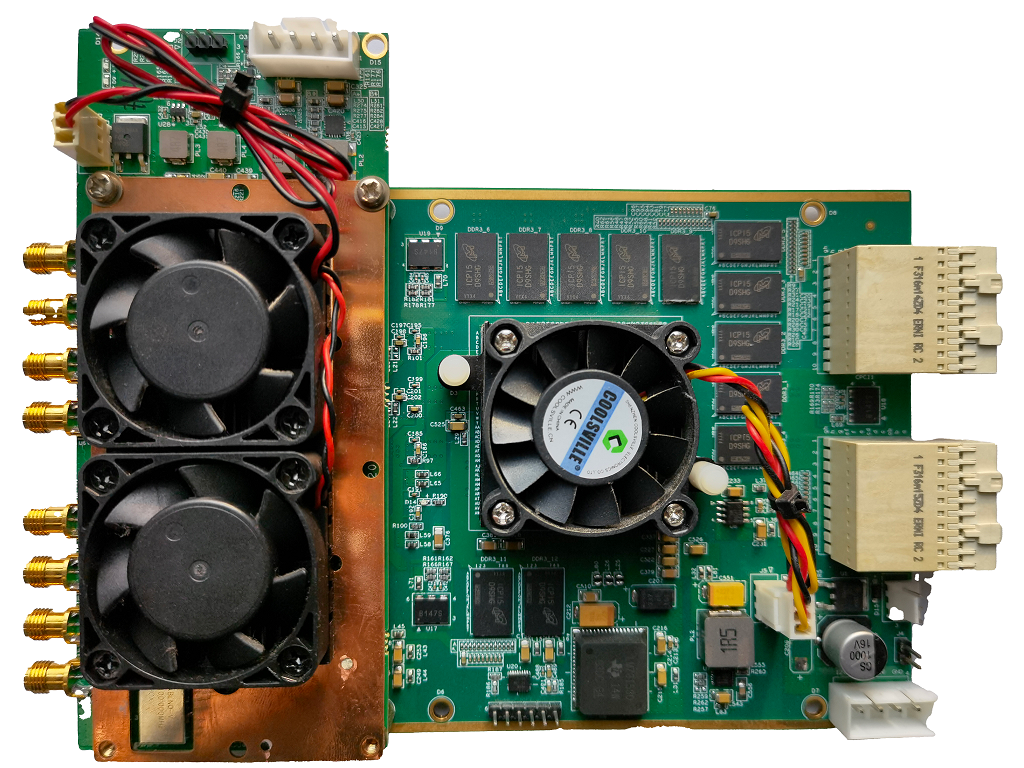}};
		\node [inner sep=0pt] (awg) at (-35mm,25mm) {\includegraphics[width=1.5in]{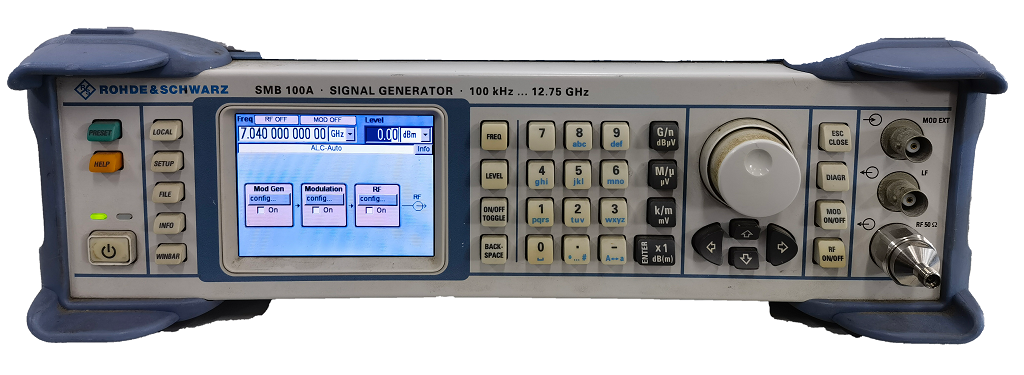}};
		\draw [red, thick, ->] (awg) |- node [above right] {} (brd);
		\callout{10mm,-25mm}{\textbf{Acquisition System}}{-6mm,-6mm}
		\callout{30mm,-25mm}{\textbf{FPGA}}{10mm,-8mm}
		\callout{10mm,25mm}{\textbf{Signal Generator}}{-18mm,25mm}
		\callout{-30mm,-25mm}{\textbf{Signal Input}}{-25mm,-10mm}
		\draw [red, thick] (-21.5mm,-18mm) rectangle (-7mm,10mm);
		\draw [red, thick] (-1.5mm,-8mm) rectangle (9mm,3.5mm);
		
		\end{tikzpicture}
		\caption{The verification platform is used to verify the anti-aliasing effectiveness of the designed SRC.}
		\label{Platform}
\end{figure}
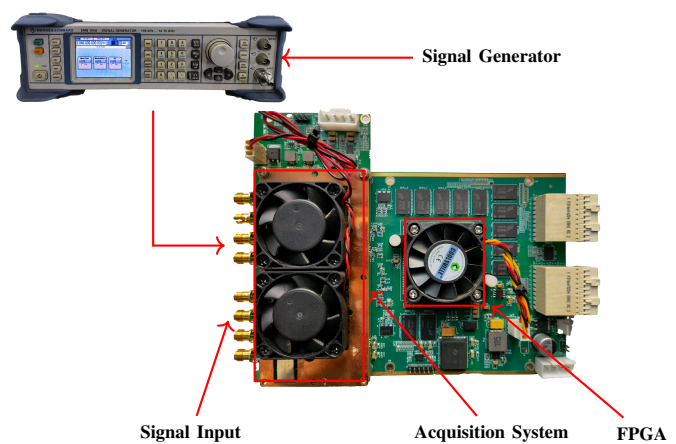

A Digital Down Converter (DDC) system with a parallel-serial SRC structure was implemented in a field programmable gate array (XCKU115-FLVA1517, Xilinx) to verify the performance of the proposed structure. The verification platform is shown in Fig. \ref{Platform}. The input radio frequency (RF) signal was firstly generated by an Signal Generator (ROHDE \& SCHWARZ, SMB100A). Two-tone signal was required in the experiment, and a radio frequency power combiner (ZFRSC-183-S+, Mini Circuit) was used to combine the signals generated by Signal Generator(SG). After that, the two-tone signal was sampled by a 20GSPS acquisition system. The quantized data passed through the mixing module of DDC to generate a baseband signal. Then the baseband signal was decimated by the SRC to reduce the sampling rate.
\subsection{Performance Analysis}
\begin{figure}[!t]
		\centering
		\subfloat[]{
			\includegraphics[width=3.5in]{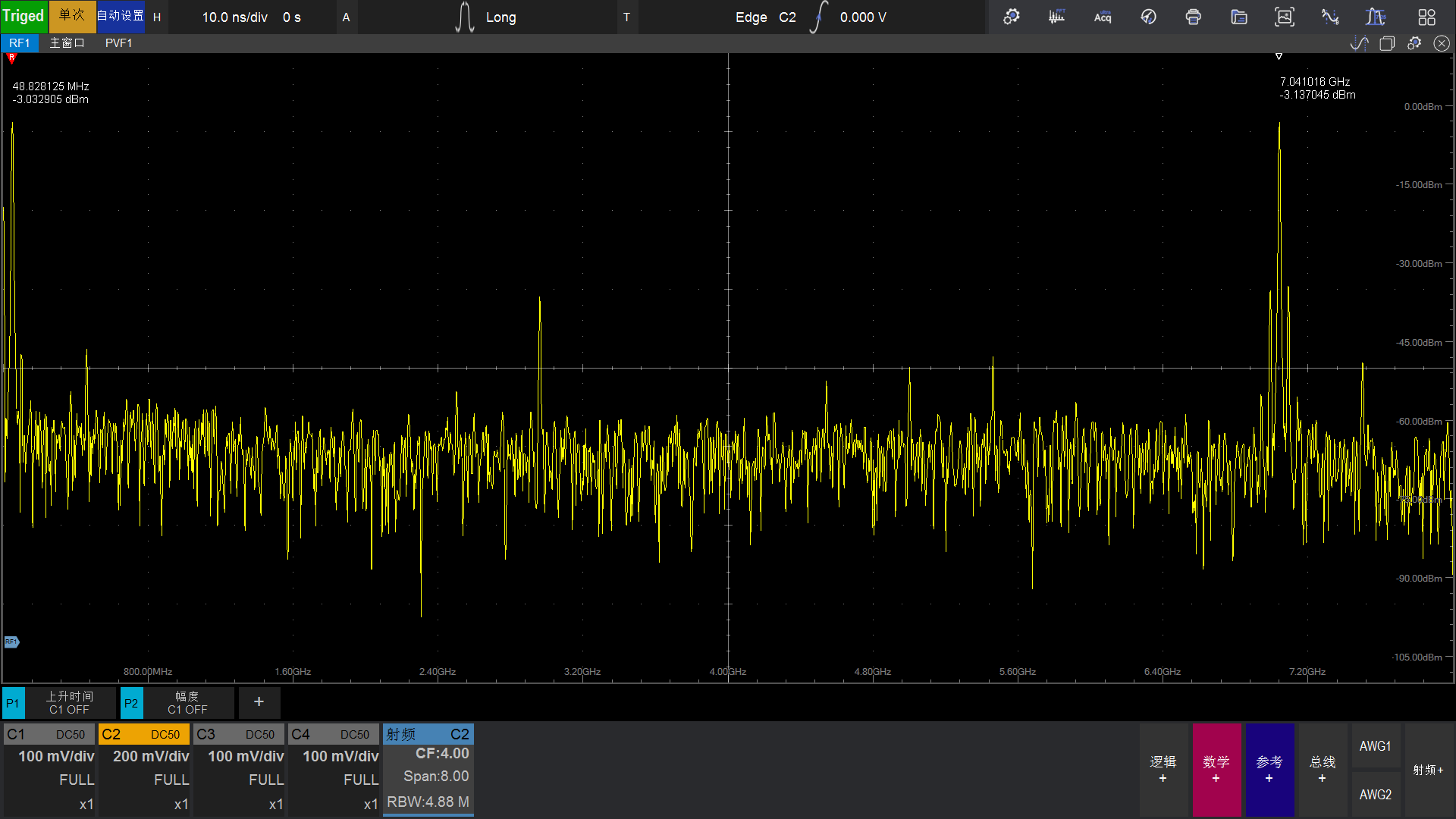}
			\label{Fig Spect1}
		}\\
		\subfloat[]{%
			\includegraphics[width=3.5in]{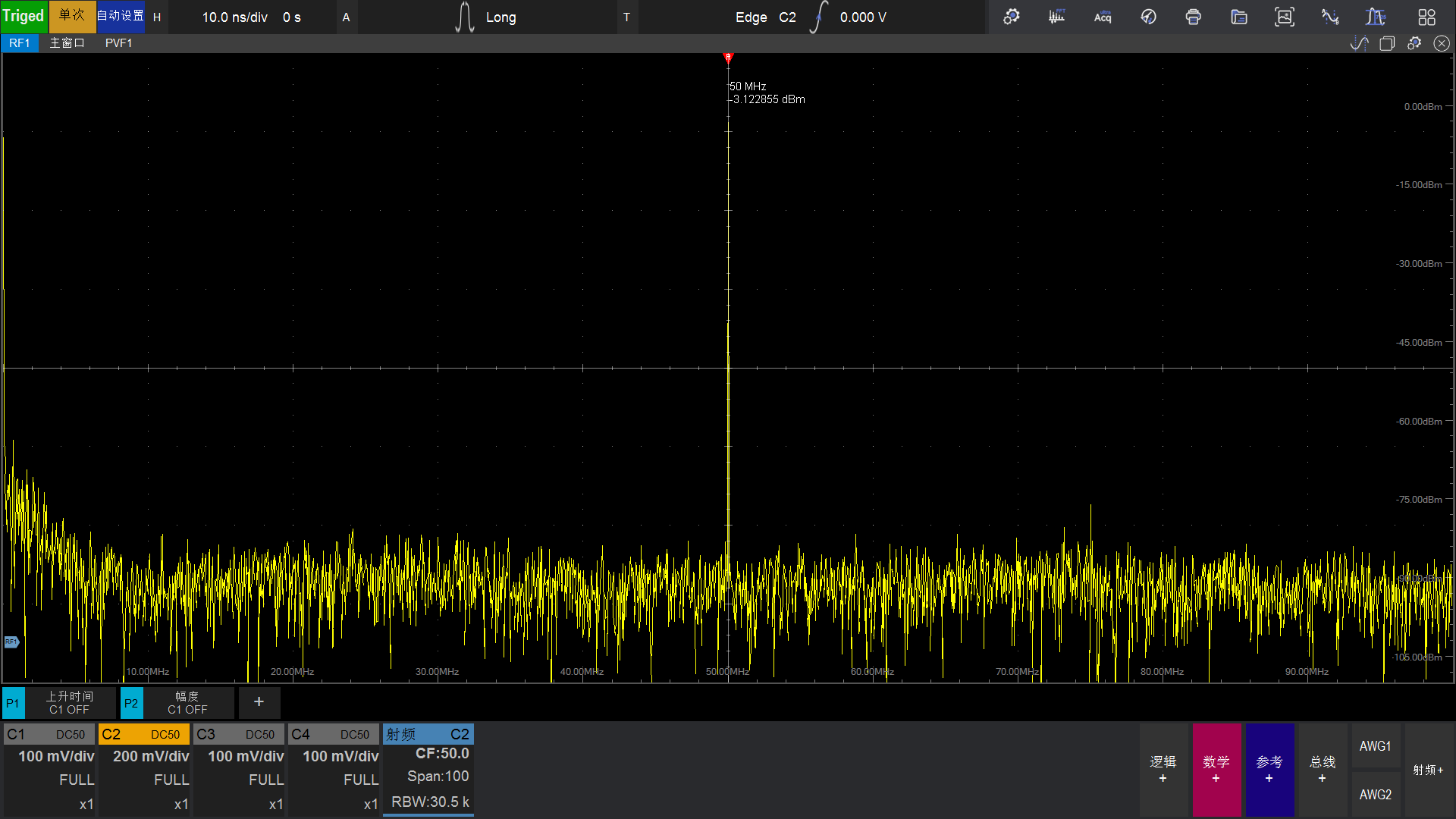}
			\label{Fig Spect2}
		}
		\caption[]{Anti-aliaing experimentation: \subref{Fig Spect1} is the spectrum of input signal. \subref{Fig Spect2} is the spectrum of decimated signal.}
		\label{Spectrum}
\end{figure}

In this experiment, the anti-aliasing performance of the designed SRC was firstly evaluated. Two Signal Generators were used to generate the desired signal and the aliasing signal. The frequency of the desired signal and aliasing signal are $50$ MHz and $7.04$ GHz respectively. The power of these two signal are all $-3$ dBm. These two signals were combined using a radio frequency power combiner to generate the two-tone signal. The spectrum of two signals is shown in Fig. \subref{Fig Spect1}, and the spectrum of the decimated signal is shown in Fig. \subref{Fig Spect2}. Note that the aliasing signal is in the noise floor after decimation, the power is nearly $-85$ dBm. That is to say, the aliasing signal had been attenuated by more than $70$ dB after decimation, which verifies the anti-aliasing performance of the proposed SRC.
\subsection{Resource Utilization Analysis}
The resource consumption of the designed SRC is shown in Table \ref{Resource}. The SRC consumes $43.4$k lookup tables (LUTs), $8.2$k flip-flops (FFs), $183$ digital signal processing(DSP) units, and has no block RAM consumption. The parallel CIC filter has the largest LUTs consumption of $37.7$k and the largest FFs consumption of $59.3$k. Considering that the lane number of the parallel CIC filter is $80$, the LUTs per input is about $542$, and the FFs per input is about $741$, obviously less than the serial CIC. The first stage of the parallel halfband filter has the largest DSP48Es consumption, but the DSP48Es per input is about $16$, still less than the serial halfband filter.

\subsection{Performance Comparison}
In this experiment, the performance of the proposed parallel-serial SRC was compared with those of previous works\cite{compare1,compare2,compare3}. The comparison result is shown in Table \ref{performance}. The proposed SRC has the highest input sampling rate($20$ GSPS), signal bandwidth($8$ GHz), and the lowest Multipliers($2$) and LUTs($542$) per input, this is achieved through the parallel CIC structure and the parallel halfband structure.  In addition, the proposed SRC has a comparable output bandwidth and decimation factor range with Liu et al. \cite{compare1}, better than the Datta et al. \cite{compare2} and Sikka et al. \cite{compare3}. The passband ripple and stopband attenuation of the proposed SRC is slightly worse than other works, this can be optimized by increasing the order of CIC filter and halfband filter. Actually, the $0.8$ dB passband ripple and $65$ dB stopband attenuation is sufficient in this SRC system.
\section{Conclusion and Future Work}
In this article, a novel parallel-serial SRC structure was proposed to process high sampling rate signals. The mathematical representation of the parallel SRC was firstly derived. The serial recursive structure of the CIC filter is transformed into a parallel recursive structure, which significantly improves the processing rate of the SRC. The zero coefficients and symmetry coefficients of the halfband filter are utilized to reduce resource consumption. Then the implementation structure of the parallel SRC was inferred based on this representation. After that, a design example of parallel-serial SRC is given. To evaluate the effectiveness of the SRC design, both performance simulation and verification experiments were conducted. The results show that the designed SRC can significantly improve the bandwidth and flexibility of SRC, and the anti-aliasing performance is sufficient. 

\section*{Acknowledgments}
The authors would like to thank
the reviewers for the time spent and their valuable comments.



 
%

\bibliographystyle{IEEEtran}
\bibliography{Reference.bib}

\newpage

%
%


%

\vfill

\end{document}